\documentclass[prl,aps,twocolumn,superscriptaddress,showpacs]{revtex4}
\usepackage[dvips]{graphicx}
\begin{document}
\author{Jing-Ling Chen}
 \email{chenjl@nankai.edu.cn}
\affiliation{Theoretical Physics Division, Chern Institute of
Mathematics, Nankai University, Tianjin 300071, P. R. China}
\author{Ming-Guang Hu}
 \affiliation{Theoretical Physics
Division, Chern Institute of Mathematics, Nankai University, Tianjin
300071, P. R. China}

\date{\today}
\title{Bell Inequality Based on Peres-Horodecki Criterion}

\begin{abstract}
We established a physically utilizable Bell inequality based on the
Peres-Horodecki criterion. The new quadratic probabilistic Bell
inequality naturally provides us a necessary and sufficient way to
test all entangled two-qubit or qubit-qutrit states including the
Werner states and the maximally entangled mixed states.

\end{abstract}

 \maketitle

One of the most striking features for quantum mechanics that differs
from classical theory is the entanglement or the nonlocality.
Arising from the EPR paradox \cite{(1935)Einstein}, the local hidden
variable theory (LHVT) was exploited by Bell and led to the
appearance of Bell inequality \cite{(1964)Bell}. The importance of
the Bell inequality is not extravagance. It is at the heart of the
study of quantum nonlocality, and makes it possible for the first
time to distinguish experimentally between a local hidden variable
model and quantum mechanics. The original Bell inequality is not
suitable for realistic experimental verification. Later on, the
Clauser-Horne-Shimony-Holt (CHSH) inequality \cite{(1969)Clauser}
was formulated, and it was a more amenable version for experimental
tests and studied the correlations between two maximally entangled
spin-1/2 particles.

For decades, quantum nonlocality has been tightly related to the
foundations of quantum mechanics, particularly to quantum
inseparability and the violation of Bell inequalities. Violation of
Bell inequalities not only tells us something fundamental about
Nature but also has practical applications. For instances, as
implied by Ekert the eavesdropping in the quantum cryptography
communication can be detected by checking the CHSH inequality
\cite{(1991)Ekert}; also Barrett \emph{et al.} have described that
testing particular nonlocal quantum correlations allows two parties
to distribute a secret key securely, the security of the scheme
stems from violation of a Bell inequality and in such a way that the
security is guaranteed by the non-signaling principle alone
\cite{(2005)Barrett}.

Despite more than four decades of active research and a vast number
of publications on the fascinating subject of Bell inequality, there
are still many questions that remain open. The CHSH inequality
simply but effectively illustrates the distinct nonlocal correlation
character of quantum world. Any entangled two-qubit pure state can
always be detected by the CHSH inequality via its violation
\cite{Gisin}. However, in the real world some states appear in pure
forms but more in mixed-state forms. In particular, for a class of
Werner states \cite{(2000)Nielson} which are used to depict the
effect of noises, there exists a range where the CHSH inequality
becomes blind \cite{Werner0,Horodecki0}. Very recently in a
significant Festschrift in honor of Abner Shimony, Gisin has
reviewed some of the many open questions about Bell inequalities
\cite{Gisin2007}. Fifteen open fundamental questions have been
listed, among which the third one is whether we can find an
inequality that is more efficient than the CHSH inequality for
testing the Werner states. Or more generally, one may ask: \emph{Is
there a universal Bell inequality, which is violated by all of the
entangled two-qubit states including the Werner states}? Such a
question seems to be some puzzling for when referring to Bell
inequality it often concerns about the obeisance of LHVT or the
violation of quantum theory, rather than the inseparability of
physical states. Yet, the increasing importance of the nonlocal
correlation characters in the Quantum Information and Communication
revolution has led us to extend the Bell inequality and test the
inseparability as well; namely, it is necessary to generalize the
original spirit of Bell inequality for distinguishing LHVT from
quantum theory to a new problem of distinguishing all separable
states from all inseparable ones.

In this Letter we show that there exists such an efficient Bell
inequality to ameliorate the above situation, and it originates
naturally from the pioneer works of Peres and Horodecki family. A
decade ago a sufficient and necessary criterion for detecting
quantum inseparability in a two-qubit or qubit-qutrit system was
presented mathematically by Peres \cite{(1996)Peres} and the
Horodecki family \cite{(1996)Horodecki}, nowadays known as the
Peres-Horodecki criterion of positivity under partial transpose (PH
criterion or PPT criterion). In 2003, Yu \emph{et al.} made a
remarkable progress that they established a three-setting Bell-type
inequality from the viewpoint of indeterminacy relation of
complementary local orthogonal observables, and proved that such an
inequality had the advantage of being a sufficient and necessary
criterion of separability with the help of PH criterion
\cite{Yu(2003)}. Since it is not easy to operate physically the
partial transpose to a subsystem, in the Letter we transform the PH
criterion into an equivalent physically utilizable Bell-inequality
form, and the new established quadratic probabilistic Bell
inequality naturally provides us a necessary and sufficient way to
test all entangled two-qubit or qubit-qutrit states including the
Werner states.

\begin{figure}
  \includegraphics[width=5cm]{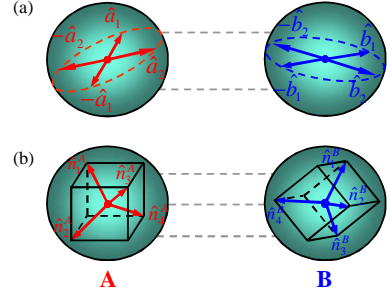}\\
\caption{\label{fig:fig1}(Color online) (a) In the Bloch sphere of a
single qubit, the four polarized unit vectors
$\{\hat{a}_1,-\hat{a}_1,\hat{a}_2,-\hat{a}_2\}$ or
$\{\hat{b}_1,-\hat{b}_1,\hat{b}_2,-\hat{b}_2\}$ employed in the
projective measurements of the CHSH-inequality lie on
  the plane embraced by a great circle; (b) the
four unit vectors $\{\hat{n}_i\}$ employed in the POVM measurements
of the PH-inequality uniformly lie on the Bloch sphere and their
endpoints occupy exactly the four vertices of a regular tetrahedron.
(Note: $\hat{n}_i^A$ and $\hat{n}_j^B$ have been rotated by $U$ and
$V$ respectively in the figure.)}
\end{figure}

Let us firstly analyze the paramount CHSH inequality from the
viewpoint of the projective measurements, and then turn to our main
result. The CHSH inequality reads
\begin{eqnarray}
   I_{CHSH}&=& \langle A_1 B_1 \rangle_\rho + \langle A_1 B_2 \rangle_\rho
+\langle A_2 B_1 \rangle_\rho-\langle A_2 B_2
\rangle_\rho\nonumber\\
    &\leq &2,
    \label{eq-Bell}
\end{eqnarray}
where $\langle A_i B_j \rangle_\rho\equiv Q_{ij}={\rm
Tr}[\rho\;(\hat{a}_i\cdot\vec{\sigma}^A)(\hat{b}_j\cdot\vec{\sigma}^B)]$
known as the so-called correlation functions, $\rho$ is the
two-qubit state shared by A and B, $\vec{\sigma}$ is the Pauli
matrix vector, $\hat{a}_1$ and $\hat{a}_2$ are the unit vectors for
the first and the second measurements performed to the subsystem A
respectively and so do $\hat{b}_1$ and $\hat{b}_2$ for the subsystem
B. According to the measurement language, the correlation functions
can be expressed in terms of joint probabilities as $\langle
A_iB_j\rangle_\rho=\sum_{m=0}^1\sum_{n=0}^1(-1)^{m+n}P(A_i=m,B_j=n)$,
with the joint probability $P(A_i=m,B_j=n)=\mathrm{Tr}[\rho\
\hat{\mathcal{P}}(A_i=m)\otimes \hat{\mathcal{P}}(B_j=n)]$, and the
projector $\hat{\mathcal{P}}(A_i=m)=\frac{1}{2}[1+(-1)^m\
\hat{a}_i\cdot \vec{\sigma}^A]$. Thus all relevant polarization
vectors $\{\hat{a}_1,-\hat{a}_1,\hat{a}_2,-\hat{a}_2\}$ and
$\{\hat{b}_1,-\hat{b}_1,\hat{b}_2,-\hat{b}_2\}$ in the Bloch spheres
of each subsystem always locate on the same plane embraced by a
great circle [see Fig. \ref{fig:fig1}(a)] so that such projective
measurements cannot acquire any information outside the plane. This
may be the reason of the invalidation of the CHSH inequality for the
whole mixed states.

To overcome this flaw, we adopt Positive Operator-Valued Measure
(POVM). An operator $E_m$ is a POVM element if it is a positive
operator satisfying $\sum_m E_m=1$ and then the complete set
$\{E_m\}$ form a POVM \cite{(2000)Nielson,Preskill}. Gisin and
Popescu have conjectured that more information is extractable if one
adopts a special class of vectors, such as $(0, 0, 1)$, $(\sqrt{8},
0, -1)/3$, $(-\sqrt{2}, \sqrt{2}, -1)/3$, $(-\sqrt{2}, -\sqrt{2},
-1)/3$, which occupy the four vertices of a regular tetrahedron
inscribed in the three-dimensional Bloch sphere \cite{Gisin(1999)}.
One may observe that these four unit vectors sum up to zero, thus it
allows us to introduce the following POVM operators:
\begin{eqnarray}
&\tilde{F}_i^A= U F_i^A {U^\dagger},\quad\quad\quad \;\;\;\;
&\tilde{F}_i^B= V F_i^B
{V^\dagger},    \nonumber\\
&F_i^A=(1+\hat{n}_i^A\cdot\vec{\sigma}^A)/4,\quad
&F_i^B=(1+\hat{n}_i^B\cdot\vec{\sigma}^B)/4,    \label{eq-POVM2}
\end{eqnarray}
where $U$ and $V$ are the general $SU(2)$ transformations for
subsystems A and B respectively, and for simplicity, the four unit
vectors $\hat{n}_i$ [see Fig. \ref{fig:fig1}(b)] that form a
tetrahedron are chosen as
\begin{eqnarray}
&& \hat{n}_1=(1,1,1)/\sqrt{3}, \; \; \;\;\;\;
\;\;\hat{n}_2=(1,-1,-1)/\sqrt{3},
\nonumber\\
&& \hat{n}_3=(-1,1,-1)/\sqrt{3}, \; \; \hat{n}_4=(-1,-1,1)/\sqrt{3}.
 \label{eq-vector2}
\end{eqnarray}
By the way, such a POVM realization has been applicable successfully
as a minimal measurement scheme for a single-qubit tomography
\cite{Rehacek(2004)}.

Accordingly, the sixteen elements $\tilde{F}_i^A\otimes
\tilde{F}_j^B$ form a POVM for the composite A-B system and $\langle
\tilde{F}^A_i \tilde{F}^B_j\rangle_\rho={\rm Tr}[\rho\;
\tilde{F}^A_i\otimes \tilde{F}^B_j] \equiv P_{ij}^{AB} $ denotes the
joint probability of the joint measurement $\tilde{F}^A_i \otimes
\tilde{F}^B_j$ on the state $\rho$. These sixteen joint
probabilities sum up to one and will be used to construct a Bell
inequality subsequently. Our main result is the following Theorem.

\emph{Theorem}: The Peres-Horodecki criterion for qubit-qubit system
is equivalent to the following quadratic Bell-type inequality:
\begin{equation}\label{ineqa-main}
I_{PH}=Y_1^2+Y_2^2-Y_3^2\leq0,
\end{equation}
where $Y_i$'s are linear combinations of the sixteen joint
probabilities $P_{ij}^{AB}$, and $I_{PH}$ denotes Bell inequality
induced from the PH criterion, alternatively one may call it the PH
inequality.

\emph{Proof}.
First we write an arbitrary projector for $AB$ system into the form
$\hat{\cal P}_{AB}=(U \otimes V) |\Phi\rangle\langle\Phi| (U\otimes
V)^\dag$, where
\begin{equation}\label{eq-Phi}
|\Phi\rangle=\sin\xi|0\rangle_A\otimes|0\rangle_B+\cos\xi|1\rangle_A\otimes|1\rangle_B,
\end{equation}
is a two-qubit pure state in the Schmidt decomposition form, the
unitary transformations $U$ and $V$ act on the parties $A$ and $B$
respectively, the angle $\xi$ is related to the Schmidt coefficient,
and $|0\rangle=(1, 0)^T$, $|1\rangle=(0, 1)^T$ are the standard
spin-1/2 bases.

Let $\rho$ be the state shared by A and B. The nonnegativity of the
density matrix $\rho$ requires that
\begin{eqnarray}\label{eq-posi-0}
\mathrm{Tr}(\rho \ \hat{\cal P}_{AB}) =\mathrm{Tr}[\rho \ (U \otimes
V) |\Phi\rangle\langle\Phi| (U\otimes V)^\dag] \ge 0.
\end{eqnarray}
On the other hand, the PH criterion states that $\rho$ is separable
if and only if its partial transpose $\rho^{T_B}$ is nonnegative,
i.e., $\mathrm{Tr}(\rho^{T_B} \hat{\cal P}_{AB}) \ge 0$, or more
generally $\mathrm{Tr}[\rho^{T_B} \; (U^A \otimes U^B) \hat{\cal
P}_{AB} (U^A \otimes U^B)^\dag]\ge 0$. By using
$\mathrm{Tr}[\rho^{T_B} \; (U^A \otimes U^B) \hat{\cal P}_{AB} (U^A
\otimes U^B)^\dag]=\mathrm{Tr}[\rho \ [(U^A \otimes U^B) \hat{\cal
P}_{AB} (U^A \otimes U^B)^\dag]^{T_B}]
= \mathrm{Tr}[\rho \ (U^AU \otimes (U^BV)^{\dag T_B})
(|\Phi\rangle\langle\Phi|)^{T_B} ((U^AU)^\dag\otimes
(U^BV)^{T_B})]$, and selecting $U^A=I$, $U^B=(V V^{T_B})^\dag$, one
arrives at an equivalent expression for the PH criterion as
\begin{eqnarray}\label{eq-posi-1}
\mathrm{Tr}[\rho \ (U \otimes V) (|\Phi\rangle\langle\Phi|)^{T_B}
(U\otimes V)^\dag] \ge 0.
\end{eqnarray}

We now combine Eqs. (\ref{eq-posi-0}) and (\ref{eq-posi-1}) together
to build the quadratic Bell inequality. With the help of
$|0\rangle_S\langle0|=1/2+\sqrt{3}(F^S_1+F^S_4-F^S_2-F^S_3)/2$, $
|0\rangle_S\langle1|=\sqrt{3}[(1+i)(F^S_1-F^S_4)+(1-i)(F^S_2-
F^S_3)]/2$, $|1\rangle_S\langle1|=1-|0\rangle\langle0|$, $
|1\rangle_S\langle0|=(|0\rangle\langle1|)^\dag$, where $S=A, B$, we
may expand $|\Phi\rangle\langle\Phi|$ in terms of POVM operators as
\begin{eqnarray}
|\Phi\rangle\langle\Phi| &= &(\sin2\xi \;\hat{X}_1 -\cos2\xi\;
\hat{X}_2 +\hat{X}_3 )/4, \label{eq-phi}
\end{eqnarray}
where $ \hat{X}_1 =
2(|0\rangle_A\langle1|\otimes|0\rangle_B\langle1|+|1\rangle_A\langle0|\otimes|1\rangle_B\langle0|)
= 6 ( F_1^A F_2^B + F_2^A F_1^B + F_3^A F_4^B + F_4^A F_3^B -F_1^A
F_3^B -F_3^A F_1^B -F_2^A F_4^B -F_4^A F_2^B)$, $ \hat{X}_2
=2(|0\rangle_A\langle0|\otimes|0\rangle_B\langle0|-|1\rangle_A\langle1|\otimes|1\rangle_B\langle1|)=
\sqrt{3} ( F_1^A +F_4^A - F_2^A - F_3^A + F_1^B +F_4^B - F_2^B -
F_3^B)$, $ \hat{X}_3 =
2(|0\rangle_A\langle0|\otimes|0\rangle_B\langle0|+|1\rangle_A\langle1|\otimes|1\rangle_B\langle1|)=1+3
( F_1^A F_1^B + F_2^A F_2^B + F_3^A F_3^B + F_4^A F_4^B + F_1^A
F_4^B + F_4^A F_1^B+F_2^A F_3^B + F_3^A F_2^B -F_1^A F_2^B - F_2^A
F_1^B-F_1^A F_3^B - F_3^A F_1^B-F_2^A F_4^B - F_4^A F_2^B-F_3^A
F_4^B - F_4^A F_3^B )$. Similarly, we have
$(|\Phi\rangle\langle\Phi|)^{T_B} = (\sin2\xi \;\hat{Y}_1
-\cos2\xi\; \hat{Y}_2 +\hat{Y}_3 )/4$, with
$\hat{Y}_i=\hat{X}_i^{T_B}$. Due to $(F_1^B)^{T_B}=1/2-F_3^B$,
$(F_2^B)^{T_B}=1/2-F_4^B$, $(F_3^B)^{T_B}=1/2-F_1^B$,
$(F_4^B)^{T_B}=1/2-F_2^B$, one may easily have $\hat{Y}_1 =
\hat{X}_1^{T_B}=6 ( F_1^A F_1^B + F_2^A F_2^B + F_3^A F_3^B + F_4^A
F_4^B -F_1^A F_4^B -F_4^A F_1^B -F_2^A F_3^B -F_3^A F_2^B)$,
$\hat{Y}_2=\hat{X}_2^{T_B}=\hat{X}_2$ and
$\hat{Y}_3=\hat{X}_3^{T_B}=\hat{X}_3$.

Substituting Eq. (\ref{eq-phi}) into Eq. (\ref{eq-posi-0}), and
using $\sin2\xi=2 t/(1+t^2)$, $\cos2\xi=(1-t^2)/(1+t^2)$ with
$t=\tan\xi$, one then gets an algebraic quadratic inequality with
respect to $t$ as $$(X_2+X_3)\ t^2+2X_1 \ t+ (X_3-X_2)\geq0,$$ where
$X_i= \mathrm{Tr}[\rho \ (U \otimes V)
 \hat{X}_i (U\otimes V)^\dag]$; since it is valid for any $t$, thus the
coefficient of $t^2$ must be nonnegative, namely the nonnegativity
of the density matrix $\rho$ ensures that $X_2+X_3\geq0$.
Similarly, Eq. (\ref{eq-posi-1}) yields $a\ t^2+b \ t+ c\geq0$,
where $a=Y_2+Y_3$, $b=2Y_1$, $c=Y_3-Y_2$, and $Y_i= \mathrm{Tr}[\rho
\ (U \otimes V)
 \hat{Y}_i (U\otimes V)^\dag]$ can be expressed in terms of the
 joint probabilities $P_{ij}^{AB}$ as: $Y_1 =
 6(P_{11}^{AB}+P_{22}^{AB}+P_{33}^{AB}+P_{44}^{AB}
 -P_{14}^{AB} -P_{41}^{AB} -P_{23}^{AB} -P_{32}^{AB})$,
 $Y_2 =
 \sqrt{3}(P_{1}^{A}+P_{4}^{A}-P_{2}^{A}+P_{3}^{A}+P_{1}^{B}+P_{4}^{B}-P_{2}^{B}+P_{3}^{B})$,
$Y_3 = 1+3(P_{11}^{AB}+P_{22}^{AB}+P_{33}^{AB}+P_{44}^{AB}
 +P_{14}^{AB} +P_{41}^{AB} +P_{23}^{AB} +P_{32}^{AB}
 -P_{12}^{AB} -P_{21}^{AB}-P_{13}^{AB} -P_{31}^{AB}
-P_{24}^{AB} -P_{42}^{AB}-P_{34}^{AB} -P_{43}^{AB} )$, here the
single probabilities satisfy $P_{i}^{A}=\sum_j P_{ij}^{AB}$ and
$P_{j}^{B}=\sum_i P_{ij}^{AB}$.  The PH criterion demands the
quadratic inequality $a\ t^2+b \ t+ c\geq0$ holds for all $t$, so
one must have (i) $a\ge 0$ and (ii) $b^2-4ac\ge 0$. The first
condition is automatically satisfied because $a=Y_2+Y_3=X_2+X_3$,
while the second condition leads to the needed quadratic Bell
inequality as shown in (\ref{ineqa-main}). This ends the proof.

The PH inequality naturally provides us a necessary and sufficient
way to test all entangled two-qubit states. To see this point
clearly we would like to provide two explicit examples as follows.
\begin{figure}
  \includegraphics[width=6cm]{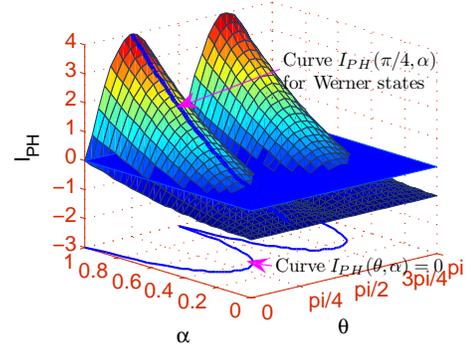}\\
\caption{\label{fig:fig2}(Color online) The maximal violation of the
PH inequality $I_{PH}$ for the general Werner state has been
plotted. Specifically, the curves for both the maximal violation of
the usual Werner state $I_{PH}(\pi/4,\alpha)$ and the boundary of
sparable states $I_{PH}(\theta,\alpha)=0$ [ i.e.,
$(1+2|\sin2\theta|)\alpha-1=0$] have been marked out (see the blue
lines).}
\end{figure}

\emph{Example 1: The Werner State.} The general two-qubit Werner
state reads
\begin{equation}\label{eq-GW}
\rho_{GW}= \alpha
|\psi(\theta)\rangle\langle\psi(\theta)|+(1-\alpha) 1\!\! 1/4,
\end{equation}
where $|\psi(\theta)\rangle=\cos\theta |00\rangle + \sin\theta
|11\rangle$, $1\!\! 1$ is a $4\times 4$ unit matrix, $\alpha
\in[0,1]$, and $\theta\in[0,\pi]$. When the parameter $\theta=\pi/4$
it reduces to the usual Werner state $\rho_W$, and when the
parameter $\alpha=1$ it reduces to the pure state
$|\psi(\theta)\rangle$. It is well known that the state $\rho_{W}$
is separable if $\alpha\le1/3$ and nonseparable if $\alpha>1/3$.
However the CHSH inequality can be violated only for the region
$\alpha\in (1/\sqrt{2}, 1]$, in other words, the Werner state
$\rho_W$ is still entangled within the region $\alpha\in (1/3,
1/\sqrt{2}]$ but the CHSH inequality fails to detect its
inseparability. For the general Werner state $\rho_{GW}$, we have
the maximum of the PH inequality as $I_{PH}^{\max}(\theta,
\alpha)=[(1+2|\sin2\theta|)\alpha-1](1+\alpha)$, see Fig.
\ref{fig:fig2}. For pure states $|\psi(\theta)\rangle$, one has
$I_{PH}^{\max}(\theta, \alpha=1)=4|\sin2\theta|$. For
$\theta=\pi/4$, $I_{PH}^{\max}(\theta=\pi/4, \alpha)=(3 \alpha
-1)(1+\alpha)$, namely, the Werner state $\rho_{W}$ is violated for
the whole nonseparable region of $\alpha\in (1/3, 1]$.

\emph{Example 2: The Maximally Entangled Mixed State.} This state
was predicted by White \emph{et al.} and had the following explicit
form \cite{White(2002)}
\begin{equation}
\rho_m=\left(\begin{array}{cccc}
g(\gamma) & 0 &0&\frac{\gamma}{2}\\
0& 1-2g(\gamma) & 0 &0\\
0&0&0&0\\
\frac{\gamma}{2}&0&0&g(\gamma)
\end{array}\right),
\end{equation}
with $g(\gamma)=\gamma/2$ for $2/3 \le \gamma \le 1$ and
$g(\gamma)=1/3$ for $0 \le \gamma<2/3$. The state is entangled for
all nonzero $\gamma$ due to its concurrence \cite{Munro(2001)}
equals to $\gamma$. It is easy to verify that the PH inequality for
such states has its maximal violation as
$I_{PH}^{\max}(\gamma)=4\gamma^2$.

The above approach can be easily generalized to a qubit-qutrit
system and one still obtains the same quadratic form of Bell
inequality as in (\ref{ineqa-main}), because the projector
$\hat{\cal P}_{AB}$ still shares the same form for arbitrary
qubit-qutrit systems. The POVM for subsystem A remains the same as
shown in Eq. (\ref{eq-POVM2}), while the POVM for subsystem B is
extended to $\tilde{F}_i^B= V F_i^B {V^\dagger}$,
$(i=1,2,\cdots,9)$, where $V$ is a general $SU(3)$ transformation,
$F^B_i=(1/9)(1+\sqrt{3}/2 \ \hat{v}_i\cdot \vec{\lambda})$,
$\vec{\lambda}=(\lambda_1, \lambda_2, \cdots, \lambda_8)$ is the
vector of $SU(3)$ Gell-Mann matrices, the factor $\sqrt{3}/2$ is
introduced to guarantee the nonnegativity, and the nine unit vectors
$\hat{v}_i$'s distribute uniformly in the eight-dimensional Bloch
space. Following the similar spirit as in the proof, one may obtain
the quadratic Bell inequality (\ref{ineqa-main}) for the
qubit-qutrit system but with different $Y_i$'s, which are linear
combinations of the $4\times 9= 36$ joint probabilities
$P_{ij}^{AB}$ of the qubit-qutrit system.

It is worthy to mention that the CHSH inequality possesses two
evident properties: (i) it is a two-setting inequality based on the
standard Bell experiment. By a standard Bell experiment, we mean one
in which each local observer is given a choice between two
dichotomic observables \cite{zukowski2,zukowski1,weinfurter,werner};
(ii) it is a linear inequality. In 2002, two research teams
independently developed Bell inequalities for two high-dimensional
systems: the first one is a Clauser-Horne type (probability)
inequality for two qutrits \cite{JLC2}; and the second one is a CHSH
type (correlation) inequality to two arbitrary $d$-dimensional
systems \cite{CGLMP}, now known as the
Collins-Gisin-Linden-Massar-Popescu (CGLMP) inequalities. The CGLMP
inequality is a two-setting inequality by the virtue of the standard
Bell experiment with possible $d$-outcomes, which includes the CHSH
inequality as a special case. The tightness of the CGLMP inequality
has been demonstrated in Ref. \cite{LM}, therefore it is impossible
to improve the CHSH inequality to be a sufficient and necessary
criterion of separability within the framework of the standard Bell
experiment. There are no physical reasons that a Bell inequality
must be linear. The PH inequality does not inherit the above two
properties and it is a quadratic four-setting inequality.

In conclusion, we have established a physically utilizable Bell
inequality based on the Peres-Horodecki criterion. The new quadratic
probabilistic Bell inequality naturally provides us a necessary and
sufficient way to test all entangled two-qubit or qubit-qutrit
states including the Werner states. The PH inequality is more
efficient than the CHSH inequality. For the crucial role of the CHSH
inequality in the previous eavesdropping detection in the Ekert's
quantum cryptography protocol, it is instructive to mention that the
PH inequality may provide a more robust approach for detecting the
eavesdropping particularly in the presence of noises. In addition,
if a Bell inequality is violated by any entangled states, such a
wisdom can be used to define the degree of entanglement $P_E$; for
two qubits, alternatively one may define $P_E={\rm Max}\{ 0,
I_{PH}^{\max}/4\}$, which is monotonic to the concurrence
\cite{Wootters(1998)}.

We thank Y. C. Liang for his valuable discussion. This work is
supported by NSF of China (Grant No. 10605013) and Program for New
Century Excellent Talents in University.

\end{document}